\documentclass[a4,aps,preprint,amsmath,amssymb,groupedaddress,nofootinbib,floatfix]{revtex4-1}

\usepackage{epsfig}
\usepackage{rotating}
\usepackage{graphicx}
\usepackage{bm}

\newcommand{\FLUKA}{{\sc fluka}}
\newcommand{\beq}{\begin{equation}}
\newcommand{\eeq}{\end{equation}}
\newcommand{\bea}{\begin{eqnarray}}
\newcommand{\eea}{\end{eqnarray}}
\newcommand{\bes}{\begin{eqnarray*}}
\newcommand{\ees}{\end{eqnarray*}}

\newcommand{\der}{\mathrm{d}}

\begin{document}

\title{\bf A search for the analogue to Cherenkov radiation by high energy
  neutrinos at superluminal speeds in ICARUS}
\author{The ICARUS Collaboration \\
M.~Antonello$^{a}$, P.~Aprili$^{a}$, B.~Baibussinov$^{b}$, M.~Baldo~Ceolin$^{b\dagger}$,
P.~Benetti$^{c}$, E.~Calligarich$^{c}$, N.~Canci$^{a}$,
F.~Carbonara$^{d}$, S.~Centro$^{b}$, A.~Cesana$^{f}$, K.~Cieslik$^{g}$,
D.~B.~Cline$^{h}$, A.~G.~Cocco$^{d}$, A.~Dabrowska$^{g}$,
D.~Dequal$^{b}$, A.~Dermenev$^{i}$, R.~Dolfini$^{c}$, C.~Farnese$^{b}$,
A.~Fava$^{b}$, A.~Ferrari$^{j}$, G.~Fiorillo$^{d}$, D.~Gibin$^{b}$,
A.~Gigli~Berzolari$^{c\dagger}$, S.~Gninenko$^{i}$,
A.~Guglielmi$^{b}$, M.~Haranczyk$^{g}$, J.~Holeczek$^{l}$,
A.~Ivashkin$^{i}$, J.~Kisiel$^{l}$, I.~Kochanek$^{l}$, J.~Lagoda$^{m}$,
S.~Mania$^{l}$,
G.~Mannocchi$^{n}$,
A.~Menegolli$^{c}$, G.~Meng$^{b}$,
C.~Montanari$^{c}$, S.~Otwinowski$^{h}$,
L.~Periale$^{n}$, 
A.~Piazzoli$^{c}$, P.~Picchi$^{n}$, F.~Pietropaolo$^{b}$,
P.~Plonski$^{o}$,
A.~Rappoldi$^{c}$,
G.~L.~Raselli$^{c}$, M.~Rossella$^{c}$, C.~Rubbia$^{a}$,$^{j}$,
P.~Sala$^{f}$, E.~Scantamburlo$^{e}$, A.~Scaramelli$^{f}$,
E.~Segreto$^{a}$, F.~Sergiampietri$^{p}$,
D.~Stefan$^{a}$,
J.~Stepaniak$^{m}$, R.~Sulej$^{m,a}$, M.~Szarska$^{g}$, M.~Terrani$^{f}$,
F.~Varanini$^{b}$, S.~Ventura$^{b}$, C.~Vignoli$^{a}$, H.~Wang$^{h}$,
X.~Yang$^{h}$, A.~Zalewska$^{g}$, K.~Zaremba$^{o}$, \\
\vspace*{2mm}
 A.~Cohen$^{q}$
\vspace*{2mm}
 }
\affiliation{
$^a$ Laboratori Nazionali del Gran Sasso dell'INFN, Assergi (AQ), Italy\\
$^b$ Dipartimento di Fisica e INFN, Universit\`a di Padova, Via Marzolo 8, I-35131 Padova, Italy\\
$^c$ Dipartimento di Fisica Nucleare e Teorica e INFN, Universit\`a di Pavia, Via Bassi 6, I-27100
Pavia, Italy\\
$^d$ Dipartimento di Scienze Fisiche, INFN e Universit\`a Federico II, Napoli, Italy\\
e Dipartimento di Fisica, Universit\`a di L'Aquila, via Vetoio Localit\`a Coppito, I-67100 L'Aquila,
Italy\\
$^f$ INFN, Sezione di Milano e Politecnico, Via Celoria 16, I-20133 Milano, Italy\\
$^g$ The Henryk Niewodniczanski, Institute of Nuclear Physics, Polish Academy of Science, Krakow,
Poland\\
$^h$ Department of Physics and Astronomy, University of California, Los Angeles, USA\\
$^i$ INR RAS, prospekt 60-letiya Oktyabrya 7a, Moscow 117312, Russia\\
$^j$ CERN, CH-1211 Geneve 23, Switzerland\\
$^l$ Institute of Physics, University of Silesia, 4 Uniwersytecka st., 40-007 Katowice, Poland\\
$^m$ National Centre for Nuclear Research, A.~Soltana 7, 05-400 Otwock/Swierk, Poland\\
$^n$ Laboratori Nazionali di Frascati (INFN), Via Fermi 40, I-00044 Frascati, Italy\\
$^o$ Institute of Radioelectronics, Warsaw University of Technology,
Nowowiejska 15/19, 00-665 Warsaw, Poland\\
$^p$ INFN, Sezione di Pisa. Largo B.~Pontecorvo, 3, I-56127 Pisa, Italy\\
$^q$ Physics Department, Boston University, Boston, Massachusetts 02215, USA\\
$^\dagger$ Deceased \\
}
\email{alfredo.ferrari@cern.ch}

\date{February 25, 2012\\
\ \ Submitted to Physics Letter B}

\begin{abstract}
  The OPERA collaboration~\cite{OperaSuper} has reported evidence of
  superluminal $\nu_\mu$ propagation between CERN and the LNGS.  Cohen
  and Glashow~\cite{cohenPRL} argued that such neutrinos should lose
  energy by producing photons and e$^+$e$^-$ pairs, through Z$^0$
  mediated processes analogous to Cherenkov radiation. In terms of the
  parameter $\delta \equiv (v_\nu^2 -v_c^2) / v_c^2$, the OPERA result
  corresponds to $\delta \approx 5\cdot 10^{-5}$. For this value\footnote{Note
    that  $(v_\nu -v_c) / v_c
    \approx \frac{\delta}{2} \approx 2.5 \cdot 10^{-5}$} of
  $\delta$ a very significant deformation of the neutrino energy
  spectrum and an abundant production of photons and e$^+$e$^-$ pairs
  should be observed at LNGS.  We present an analysis based on the
  2010 and part of the 2011 data sets from the ICARUS experiment,
  located at Gran Sasso National Laboratory and using the same 
  neutrino beam from CERN.
  We find that the rates and deposited energy distributions of neutrino
  events in ICARUS agree with the expectations for an unperturbed spectrum
  of the CERN neutrino beam.  Our results therefore refute a superluminal
  interpretation of the OPERA result according to the Cohen and
  Glashow prediction~\cite{cohenPRL} for a weak current analog to
  Cherenkov radiation.  In particular no superluminal Cherenkov like
  e$^+$e$^-$­pair or $\gamma$ emission event has been directly
  observed inside the fiducial volume of the “bubble chamber like”
  ICARUS TPC-‐LAr detector, setting the much stricter limit of $\delta
  < 2.5\cdot 10^{-8}$ at the 90\% confidence level, comparable with
  the one due to the observations from the SN1987a~\cite{SN1987a}.
\end{abstract}

\maketitle

\section{Introduction \label{Sect:intro}}
The OPERA collaboration has presented evidence of superluminal
neutrino propagation~\cite{OperaSuper}, reporting a travel time
between CERN and the LNGS laboratory some 60~ns shorter than expected
for travel at light speed.  The OPERA result corresponds to $\delta
\equiv (v_\nu^2 -v_c^2) / v_c^2 \approx 5\cdot 10^{-5}$ with only
small variations over the energy domain of the detected neutrinos.
Observations of neutrinos from Supernova SN1987a at much lower
energies around 10 MeV yield a strong constraint~\cite{SN1987a}
$\delta < 4\cdot 10^{-9}$ implying a rapid increase with energy of the
hereby alleged anomaly.  

As is well known, charged particles travelling at speeds exceeding
that of light emit characteristic electromagnetic radiation known as
Cherenkov radiation.  Because neutrinos are electrically neutral,
conventional Cherenkov radiation of superluminal neutrinos does not
arise or is otherwise weakened. However neutrinos do carry electroweak
charge and, as pointed out by Cohen and Glashow~\cite{cohenPRL}, may
emit Cherenkov-like radiation via weak interactions when traveling at
superluminal speeds.  Cohen and Glashow argue that, under the
assumptions of the usual linear conservation of energy and momentum
and only slow variation of $\delta$ over the OPERA-relevant energy
domain, superluminal neutrinos would radiate and lose energy via the
three following processes
\bea
\nu_x &\rightarrow& \nu_x + \gamma \label{eq:nugamma} \\
\nu_x &\rightarrow& \nu_x + \nu_y + \bar{\nu}_y \label{eq:nununu} \\
\nu_x &\rightarrow& \nu_x + e^+ + e^- \label{eq:nupair} 
\eea 
The emission rate and energy loss is dominated by the third process,
which is kinematically allowed under the stated assumptions.  The
process~\ref{eq:nupair}, from now on referred to as pair
bremsstrahlung~\cite{cohenPRL}, proceeds through the neutral current
weak interaction and has a threshold energy $E_{thr} \approx 2
m_e/\sqrt{\delta}$ corresponding to about 140~MeV for the OPERA value
of $\delta$.  In the high energy limit the electron and neutrino
masses may be neglected, and Cohen and Glashow~\cite{cohenPRL}
compute\footnote{These expressions have corrected a numerical factor
  error in~\protect\cite{cohenPRL} of $4/5$.} the rate of pair emission
$\Gamma$, and the associated neutrino energy loss rate $dE/dx$ to
leading order in $\delta$:
\bea
\Gamma &=& \frac{2}{35} \frac{G_F^2}{192 \pi^3} E_\nu^5 
                      \delta^3 \label{eq:gamma}\\
\frac{\der E}{\der x} &=& -\frac{5}{112} \frac{G_F^2}{192 \pi^3}
E_\nu^6 \delta^3
\label{eq:dedx}
\eea
Note that the average fractional energy loss per pair emission event
is $\frac{dE /dx}{\Gamma E} \approx 0.78$; that is, about
$\frac{3}{4}$ of the neutrino energy is lost on average with each
emission.  Furthermore, under the approximation of a continuous energy
loss, the integration of $dE/dx$ over a distance $L$ provides the
following result for the final neutrino energy, $E_{\nu f}$, as a
function of the initial energy, $E_{\nu i}$: 
\beq \frac{1}{E_{\nu f}^5} - \frac{1}{E_{\nu i}^5} = \frac{5}{112}\frac{G_F^2}{192
  \pi^3} \delta^3 L \label{eq:efin} 
\eeq
Folding the initial neutrino spectrum of the CERN to Gran Sasso
neutrino beam with the energy at Gran Sasso predicted with the above
formula, the expected neutrino interaction rates and pair
bremsstrahlung rates as a function of $\delta$ may be estimated. In
particular, for $\delta = 5\cdot 10^{-5}$ and $L = 732$~km,
equation~\ref{eq:efin} would predict that few neutrinos with energy
larger than $\approx$13~GeV would reach Gran Sasso.

However, since neutrinos lose a large fraction of their energy at each
pair creation event, and the resulting deflection angles are not
negligible with respect to the angular width of the CNGS neutrino
beam, a continuous energy loss approximation is suitable only for a
qualitative estimate of the spectral distortion.

A full three-dimensional Monte~Carlo calculation of the propagation of
neutrinos from CERN to Gran Sasso has therefore been performed for
several values of $\delta$ (see section~\ref{Sect:Simul}).  As a
result, the expected rates of neutrino charged current interactions
and of e$^+$e$^-$ pair events have been obtained as a function of
$\delta$ and are presented in the next section.

The rates obtained in this way have been compared with the results of
year 2010 and part of the year 2011 exposures of the reconstructed
neutrino charged current events in the ICARUS/CNGS2 experiment located
in Hall~B of the Gran Sasso Laboratory.
\section{Simulation results \label{Sect:Simul}}
A full 3-dimensional simulation of the generation and transport of CNGS
neutrinos from CERN to Gran Sasso while undergoing pair bremsstrahlung has
been performed using the official CNGS simulation
setup~\cite{CNGSsim1,CNGSsim2}, based on the \FLUKA~\cite{FLUKA1,FLUKA2} 
Monte~Carlo transport code.

Accounting for the threshold, and under the hypothesis that
$\delta$ does not vary significantly in the range of energies of interest, 
the pair bremsstrahlung interaction rate differential in the neutrino 
energy loss, $w$, and in the pair invariant mass $s_{e^+e^-}\equiv s
\delta$, can be expressed as:
\bea
\frac{\der^2 \Gamma}{\der w \der s} &=& \frac{G_F^2 \delta^3}{192 \pi^3}
\frac{s}{E_\nu^2} \left({1 - \frac{s_0}{s}}\right)^{\frac{3}{2}}
\left[{2 E_\nu (E_\nu - w) + w^2 - s}\right] \nonumber \\
& \ & \label{eq:d2gdwds} \\
s_0 &\equiv& \frac{4 m_e^2}{\delta}
\label{eq:s0def}
\eea
The kinematical limits are given by:
\beq
E^2_\nu > w^2 > s > s_0
\label{eq:limits}
\eeq
The neutrino deflection angle $\Psi$ with respect to the incident neutrino
direction can be expressed as:
\beq
\cos{\Psi}   = 1 - \delta \frac{w^2-s(1+\delta)}{2 E_\nu (E_\nu - w)}
\label{eq:psi}
\eeq
and the e$^+$e$^-$ pair angle as:
\beq
\cos{\theta} = 1 - \frac{\delta}{2}\left(1-\frac{w}{E_\nu}\right)
\left(1-\frac{s}{w^2}\right) \label{eq:theta}
\eeq
The resulting mean free path for a 19~GeV neutrino 
(the fluence-averaged energy of CNGS neutrinos) 
is $\approx 490$~km 
for $\delta=5\cdot 10^{-5}$, and the
deflection angle is of the order of $\sqrt{\delta}$, comparable with
the angular width of the neutrino beam. Hence the need for a
full Monte~Carlo simulation of the neutrino propagation to Gran Sasso.
\begin{figure}[hbt!]
\begin{center}
\includegraphics[bb=10 80 560 580, width=1.\linewidth]{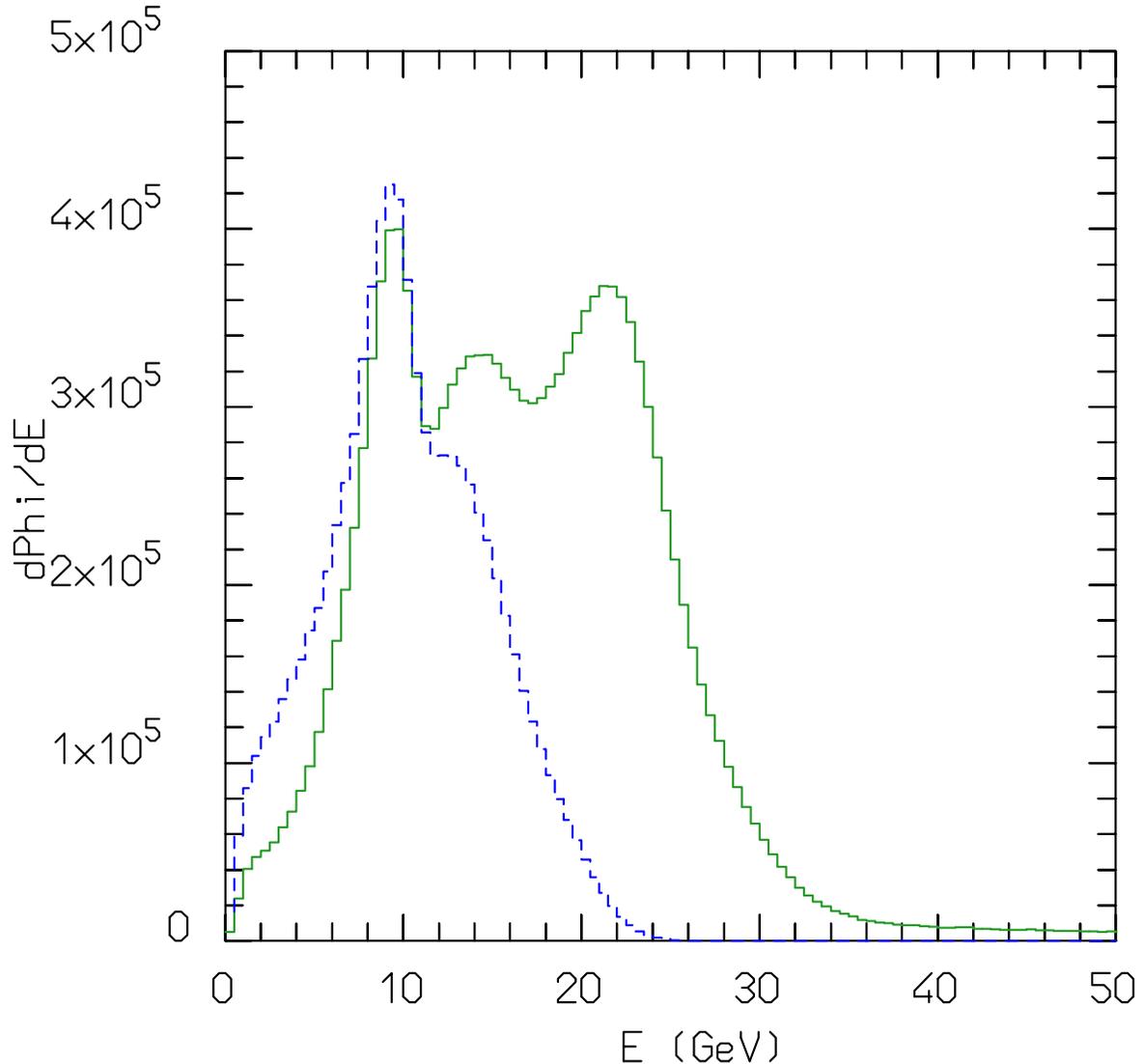}
\caption{Computed $\nu_\mu$ spectra at Gran Sasso for $\delta=0$ (solid
  line, green), and for $\delta=5\cdot 10^{-5}$ (dashed line, blue). The
  units are $\nu\ cm^{-2}\ GeV^{-1}\ 10^{-19} pot^{-1}$, 
  where pot=protons on target.
\label{fig:nuspect}}
\end{center}
\end{figure}
\begin{figure}[hbt]
\begin{center}
\includegraphics[bb=10 80 560 580, width=1.\linewidth]{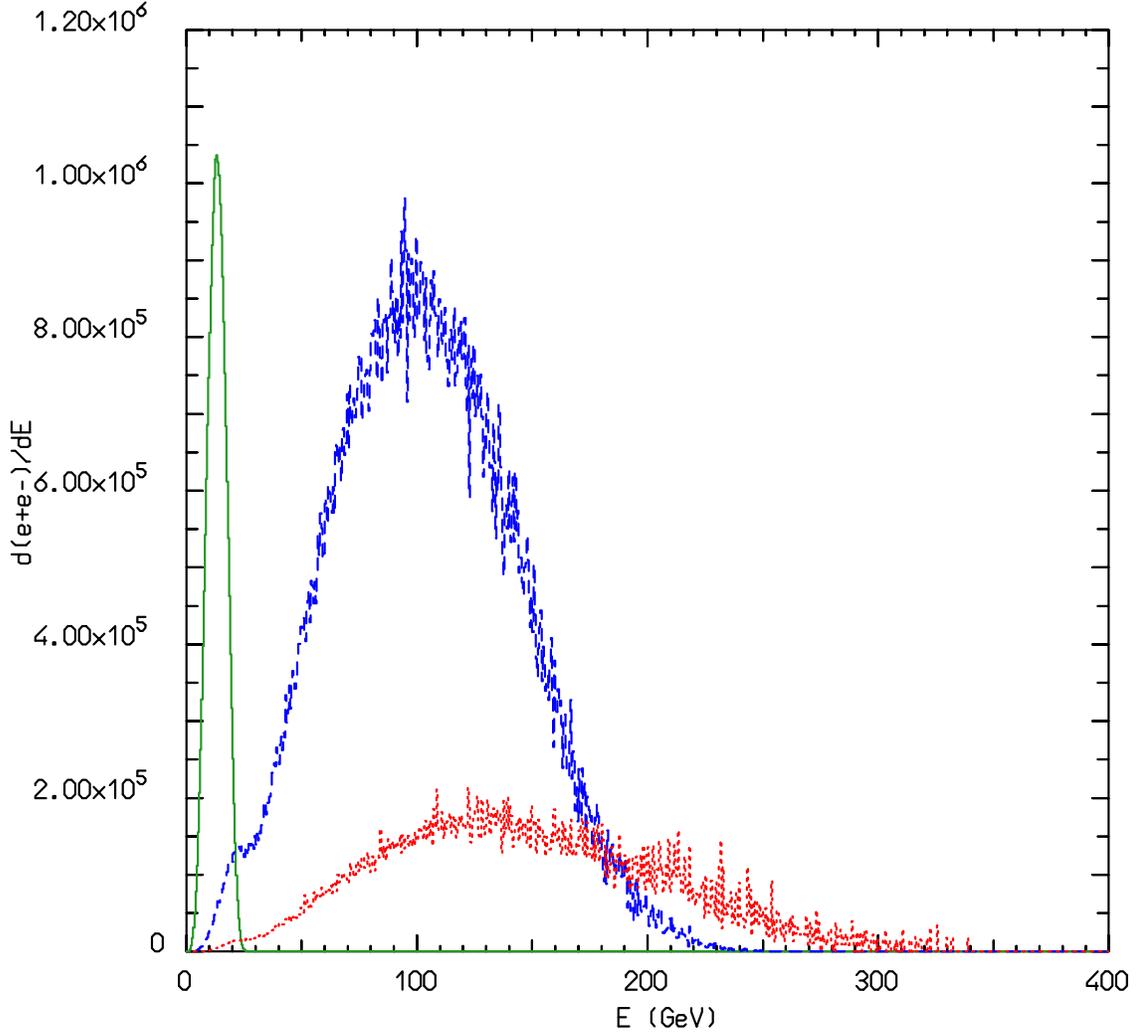}
\caption{Computed e$^+$e$^-$ pair spectra at Gran~Sasso for 
   $\delta=5\cdot 10^{-5}$ (solid
   line, green), $\delta=1\cdot 10^{-6}$ (dashed line, blue, multiplied by
   1000), and $\delta=5\cdot 10^{-8}$ (dotted line, red, multiplied by
   $10^6$). The event rate units are GeV$^{-1}$ for a 1~kt detector and
   10$^{19}$ protons on target (pot).
\label{fig:pairspect}}
\end{center}
\end{figure}
\begin{figure}[hbt]
\begin{center}
\includegraphics[bb=1 1 570 560, width=1.\linewidth]{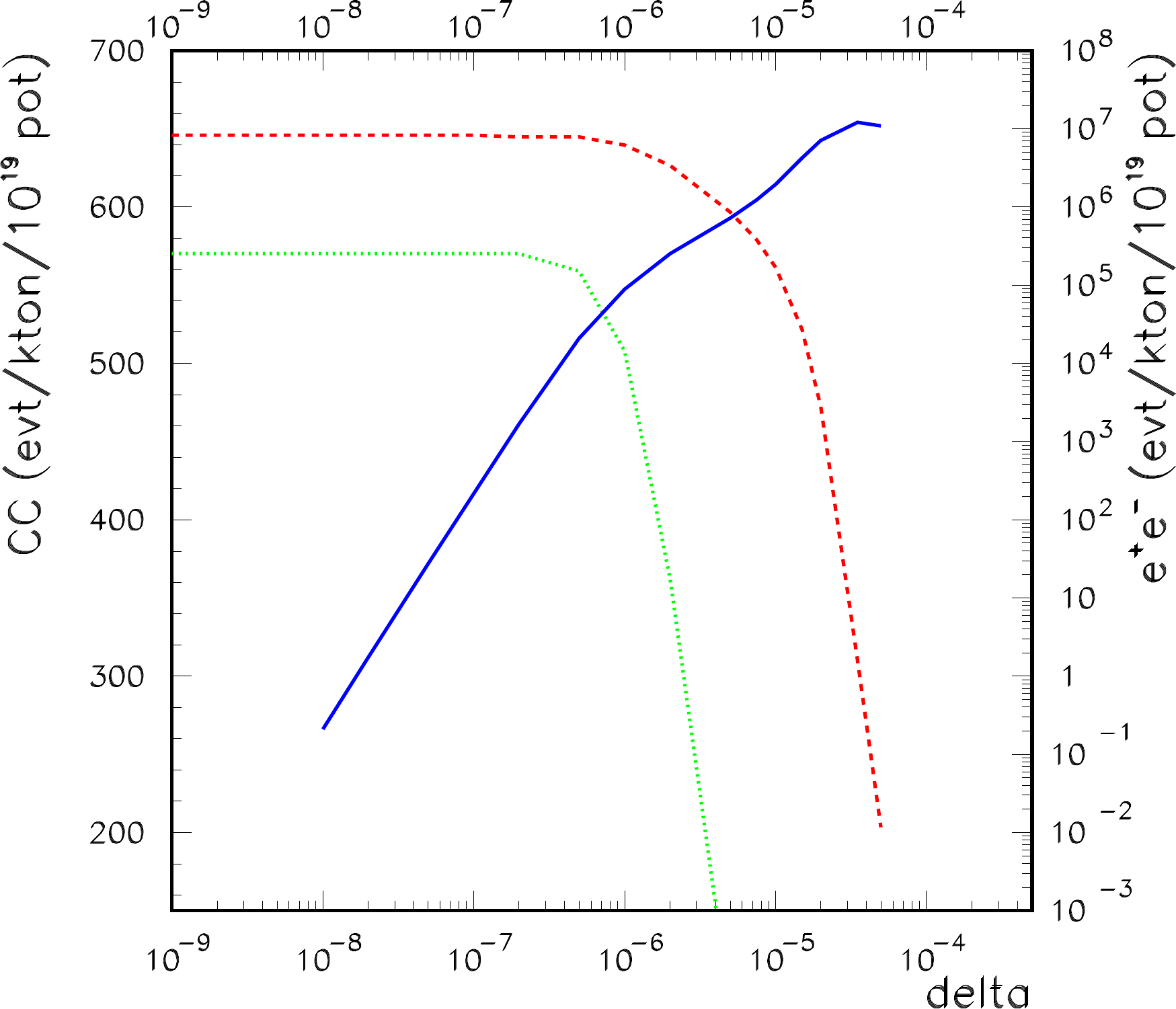}
\caption{Computed neutrino charged current rates (dashed line, red, left
  vertical scale),
   $\nu_\mu$ and $\bar\nu_\mu$ charged current rates (x 100), E$_\nu>$~60~GeV
  (dotted line, green, left vertical scale), e$^+$e$^-$ rates (solid line,
  blue, right vertical scale) at Gran Sasso, for $10^{19}$ protons on target
  (pot) and 1~kt detector.
\label{fig:allrates}}
\end{center}
\end{figure}
\begin{table}
\caption{Expected neutrino and e$^+$e$^-$ rates at Gran~Sasso. All rates
  are given for a 1~kt detector and $10^{19}$~pot.
\label{tab:rates}}
\begin{ruledtabular}
\begin{tabular}{| c | c | c | c | c |}
                &    CC   &   NC   & CC$>$60 GeV  & e$^+$e$^-$ \\
$\delta$        & (all flavours) & (all flavours) & ($\nu_\mu+\bar\nu_\mu$) & \\
\hline
0               & 644     &  203   &  57   &    0       \\
$5\cdot 10^{-8}$ & 644     &  203   &  57   &   27             \\
$5\cdot 10^{-7}$ & 643     &  203   &  56   & $2.1 \cdot 10^4$  \\
$5\cdot 10^{-6}$ & 594     &  188   &  8.5  & $7.2 \cdot 10^5$  \\
$5\cdot 10^{-5}$ & 203     &   85   & $<10^{-6}$ & $1.1 \cdot 10^7$  \\
\end{tabular}
\end{ruledtabular}
\end{table}

All results presented in this section are for $10^{19}$ protons on target
(pot) and, for rates, for a detector (Argon) mass of 1~kt. In this way they can
be easily rescaled to whichever Gran Sasso detector mass and exposure,
neglecting the minor differences in neutrino cross sections in an energy
range dominated by DIS among Argon and other materials. The nominal
yearly number of protons for CNGS is $4.5 \cdot 10^{19}$~pot.
The statistical error on integrated values (e.g. total rates, total fluence,
etc) is less than one percent in all cases. The systematic error on the
computed neutrino (and hence e$^+$e$^-$ pairs) rates is mostly due to the
uncertainties in the hadron production model of \FLUKA, and can be
conservatively estimated to be lower than 10\% (see for 
example~\cite{Collaz,FLUKA1}).

The unperturbed ($\delta=0$) fluence spectra of CNGS $\nu_\mu$ at Gran
Sasso, and the one
computed corresponding to $\delta = 5\cdot 10^{-5}$ are shown in 
Fig.~\ref{fig:nuspect}: the lack in the latter spectrum of the sharp 12.5~GeV 
ridge predicted by formula~\ref{eq:efin} can be easily appreciated. 

The computed spectra of the expected events due to e$^+$e$^-$ pairs at 
Gran Sasso are shown in Fig.~\ref{fig:pairspect}, for
$\delta~=~5~\cdot~10^{-5},\ 1~\cdot~10^{-6},\ 5~\cdot~10^{-8}$ respectively. 

The computed (anti)neutrino charged and neutral current rates 
(all flavours included), the
charged current rates for $\nu_\mu$ and $\bar\nu_\mu$ with energy above
60~GeV, and the pair bremsstrahlung rates at Gran Sasso are presented in
Fig.~\ref{fig:allrates}. The expected rates are summarized in
Table~\ref{tab:rates} for a few representative values of $\delta$.
\section{Experimental results and related constraints \label{Sect:Exp}}
\begin{figure}[hbt!]
\begin{center}
\includegraphics[bb=1 1 822 512, width=1.\linewidth]{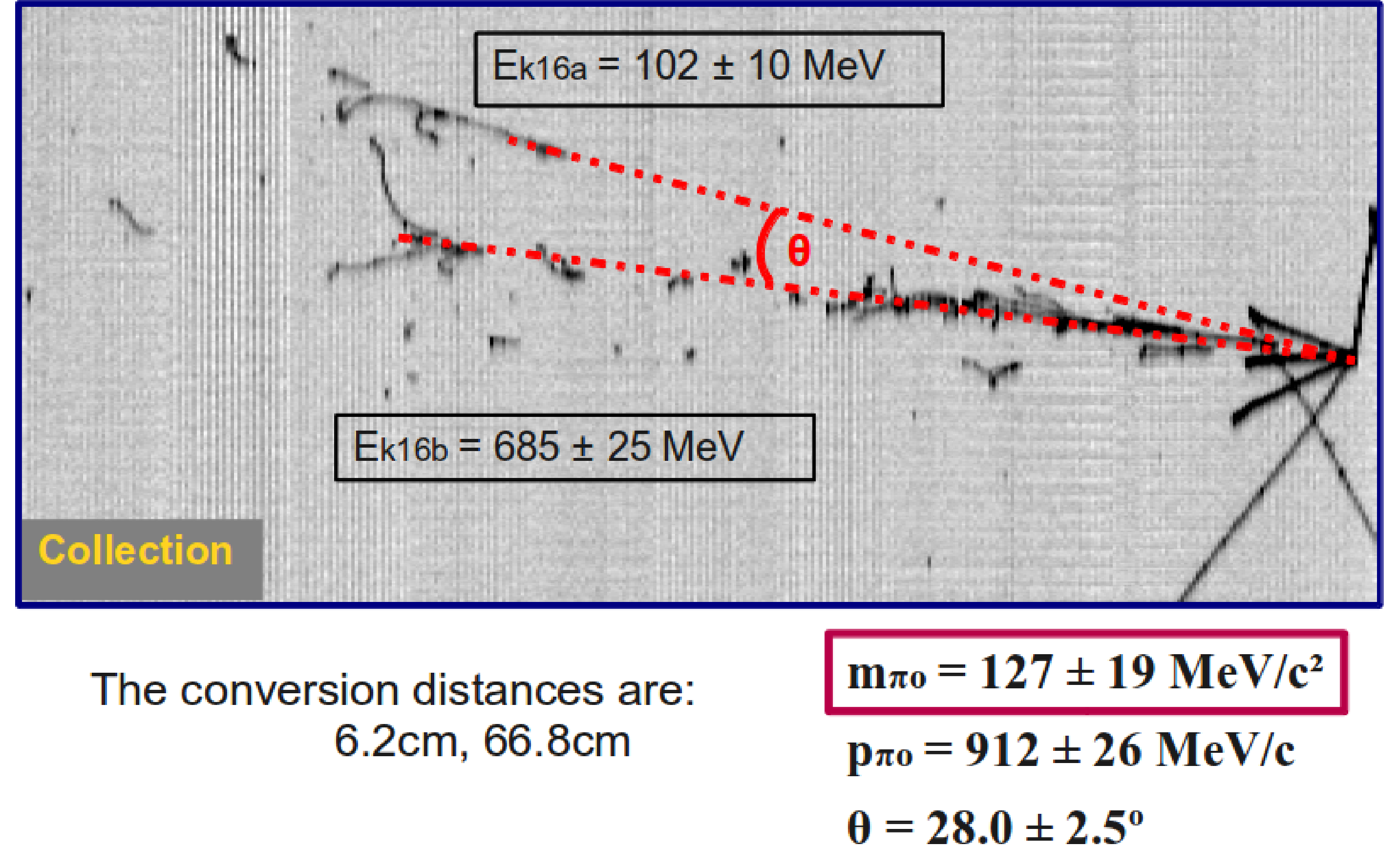}
\caption{Typical event recorded in ICARUS.  Evidence for a pair
of $\gamma$'s from a $\pi^0$ (tracks 16a and 16b) with a momentum of 
912~MeV/c pointing
at the primary vertex, showing the typical behavior of $\gamma$ 
conversions in the TPC-LAr Imaging chamber.
\label{fig:pi0}}
\end{center}
\end{figure}
\begin{figure}[hbt!]
\begin{center}
\includegraphics[bb=10 10 485 485, width=1.\linewidth]{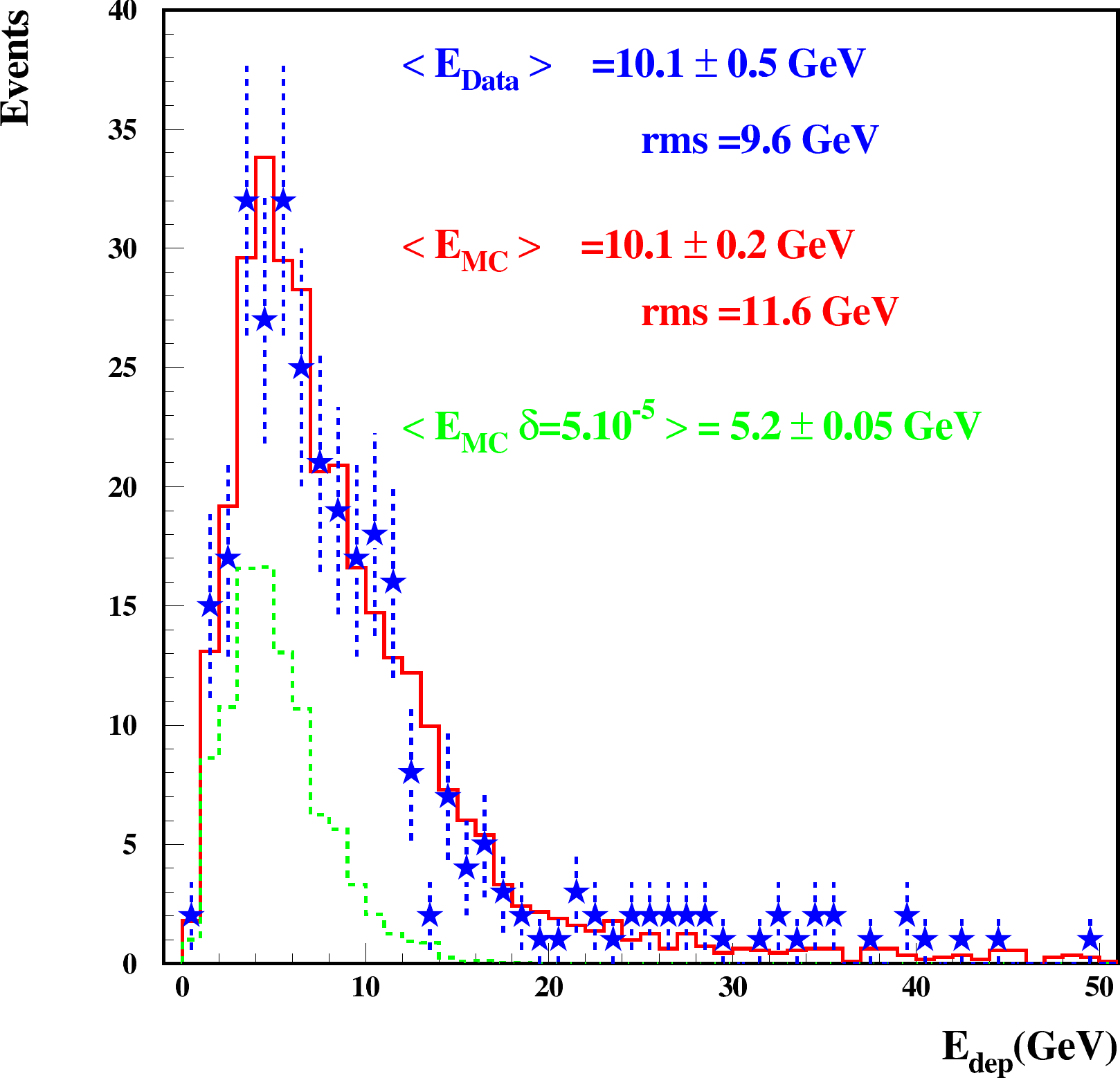}
\caption{
Experimental raw energy E$_{dep}$ distribution for $\nu_\mu$ and
$\bar\nu_\mu$ CC interactions in ICARUS
 (blue symbols) compared with the
Monte~Carlo expectations for an unperturbed spectrum (red solid histogram), and
for $\delta=5\cdot 10^{-5}$ (green dashed histogram).
\label{fig:ehadr}}
\end{center}
\end{figure}
%
%
\begin{figure}[hbt!]
\begin{center}
\includegraphics[bb=10 10 485 485, width=1.\linewidth]{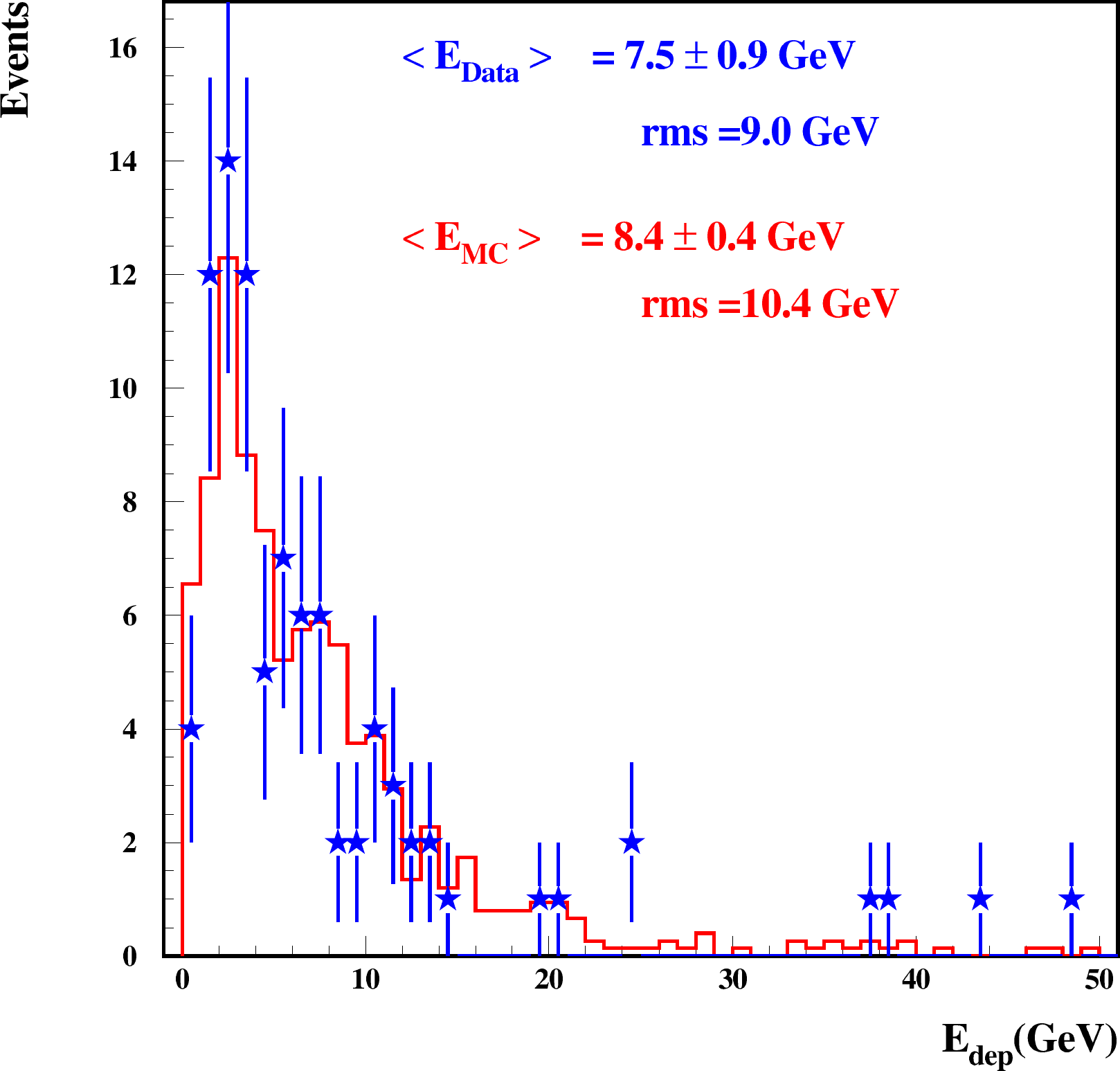}
\caption{Experimental (blue symbols) raw energy deposition distribution 
for neutral current events compared with the Monte~Carlo
expectations (red histogram) for the unperturbed CNGS spectrum. Only
experimental and Monte~Carlo events with energy deposition greater than
500~MeV have been considered.
\label{fig:ncedep}}
\end{center}
\end{figure}
The ICARUS experiment~\cite{ICARUS0,ICARUS1,ICARUS2,ICARUS3} consists of 760~t
of super-pure Liquid Argon operated as a very high resolution Time
Projection Chamber, namely a Bubble Chamber--like detector recording all
events with an energy deposition in excess of a few hundred MeV within a
window of 60~$\mu$s centered around the neutrino pulse from the CERN-SPS.
The hadronic and electromagnetic energy depositions of each event are
accurately measured by calorimetric determination while the muon momenta
are measured with the help of the multiple scattering along the very many
points of the long muon tracks.  
An example of an event with a pair of $\gamma$'s
produced by a secondary $\pi^0$ is shown in Fig.~\ref{fig:pi0}.

In the following an analysis of neutrino interactions from the 
2010 and part of the 2011 CNGS runs is compared with the expectations for 
different values of $\delta$. While the fine details of the analysis could 
still somewhat evolve, it will be shown that $\delta$ values of the order 
of the one claimed by OPERA can be readily excluded on the basis of the 
observed rates and raw energy deposition spectra. In order to carry out as
much as possible a bias-free analysis, the raw energy deposition
distributions as recorded in ICARUS with the calorimetric methods
have been compared with the expectations of a full Monte~Carlo 
simulations of the detector: only the correction for the signal
quenching in LAr has been applied to both the experimental and Monte~Carlo
results.

A dedicated search for e$^+$e$^-$ events has been also
carried out using the same exposure. This analysis constrains 
$\delta$ to values a few order of magnitude
smaller than the one claimed by OPERA.

The number of collected neutrino interactions has been compared with the
predictions for the CERN SPS neutrino beam in the whole energy range,
corrected for the fiducial volume and DAQ dead‐time. The experimental
analysis
corresponds to an integrated exposure of $6.70 \cdot 10^{21}$~t$\cdot$pot.
This exposure is the combination of a fiducial volume of 447~t
of Liquid Argon and a total number of
protons on target (pot) at CERN of $4.9 \cdot 10^{18}$ for the year 2010, and
434~t and $1.04 \cdot 10^{19}$~pot for the fraction of the year 2011
analyzed up to now. This exposure applies fully to the search for pair 
bremsstrahlung events which do not require any further cut on the 
fiducial volume because of their expected clean signature.

In order to identify $\nu_\mu$ and $\bar\nu_\mu$ charged current (CC) 
as well as neutral curent (NC) events, further cuts have been applied 
to the fiducial volume: in particular events with the vertex in the last
2.5 metres of the detector have not been considered for this analysis in
order to identify cleanly possible muon tracks.
The total number of identified $\nu_\mu$ and $\bar\nu_\mu$ CC events, and of 
neutral current events, are compared in Table~\ref{tab:icarates}: 
21 events cannot be safely assigned despite the reduced fiducial volume. 
The resulting reference exposure for NC and CC events is 
$5.05 \cdot 10^{21}$~t$\cdot$pot.

\begin{table}
\caption{Observed and expected neutrino and e$^+$e$^-$ rates at Gran~Sasso
  for the ICARUS experiment. Both the experimental and computed rates
  are normalized to the exposure used for the analysis of the 
  corresponding experimental channel (see text for details).
\label{tab:icarates}}
\begin{ruledtabular}
\begin{tabular}{| c | c | c | c |}
  Rates         &  Observed & Expected   & Expected \\
                &           & $\delta=0$ & $\delta=5\cdot 10^{-5}$ \\
\hline
 CC                   & 308    &  315$\pm$5   &  98.1$\pm$2 \\
 NC                   &  89    &  93.1$\pm$3   &  33.0$\pm$1  \\
 $\nu_\mu$ CC, $E_{dep} > 25$~GeV &  25     &  18 $\pm$1.3  &  $< 10^{-6}$ \\
 e$^+$e$^-$           &   0    &   0   &  $7.4 \cdot 10^6$  \\
\end{tabular}
\end{ruledtabular}
\end{table}

The measured raw energy deposition E$_{dep}$ for $\nu_\mu$ and 
$\bar\nu_\mu$ CC  events
as obtained from a calorimetric measurements corrected only for signal
quenching is presented in Fig.~\ref{fig:ehadr}.
The experimental distribution is compared with a full Monte~Carlo
simulation of the experimental apparatus for $\delta=0$ and 
$\delta=5\cdot 10^{-5}$. The experimental finding matches well 
the Monte~Carlo expectations for the unaffected CERN neutrino beam.

The analysis of neutral current events gives similar results. 
The experimental and expected spectra of the deposited energy for NC events
are presented in Fig.~\ref{fig:ncedep}: only events with experimental or
Monte~Carlo energy deposition in excess of 500~MeV have been considered,
in order to avoid possible misidentifications or inefficiencies.

The strong constraints of Cohen and Glashow~\cite{cohenPRL} 
predict that a superluminal
high energy neutrino spectrum will be heavily depleted and distorted after
$L =732$~km from CERN to LNGS: in particular for $\delta$ in the range 
indicated by
OPERA the charged current rate would be reduced to roughly 32\% of the
expected one, the average $\nu_\mu$ energy would be 12.1~GeV (against
19~GeV), the average energy of $\nu_\mu$ undergoing charged current 
interactions would be 12.5~GeV (against 28.7~GeV), and no neutrino
interactions should be observed above 30~GeV.
Indeed at $\delta = 5\cdot 10^{−5}$ essentially no 
E$_\nu >$~30~GeV should arrive from CERN to LNGS, while our results are 
indicating no visible deviation of the incoming neutrino beam with respect 
to the expected rate and energy distribution.  This result confirms the
inconsistence between the OPERA $\delta$ value and the observed neutrino 
rate and spectrum already reported in ref.~\cite{cohenPRL}.

In addition, with ICARUS — a bubble chamber like detector — a much more
stringent limit to $\delta$ may be set from the direct observation inside the
ICARUS detector volume of Cherenkov like events (eq.~\ref{eq:nugamma},\ref{eq:nupair})
generated by the passing superluminal neutrinos.  These events
would be characterized either by a single gamma ray converting
into an e$^+$e$^-$ pair
(eq.~\ref{eq:nugamma}) and/or two single electrons (eq.~\ref{eq:nupair}) 
both with no hadronic recoils in the incoming neutrino direction.  
The transverse momenta of the particles in
the events (\ref{eq:nugamma}) and (\ref{eq:nupair}), as determined 
by the centre of mass system, are
however far too small to be experimentally observable.  Therefore events of
both types (\ref{eq:nugamma}) and (\ref{eq:nupair}) would appear as narrow 
e$^+$ e$^-$ pairs pointing
directly to the beam direction, with no detectable hadronic activity.  
The rate of such events for the ICARUS detector exposure under
consideration ($6.70 \cdot 10^{21}$~t$\cdot$pot) can be derived from 
those presented in Fig.~\ref{fig:allrates}.
With the OPERA result ($\delta \approx 5 \cdot 10^{-5}$)
more than $7 \cdot 10^6$ electron positron pairs should have been 
observed for this exposure, each with an energy spectrum peaked around
10~GeV (see Fig.~\ref{fig:pairspect}).  This number
should not be a surprise since the CC neutrino event rates of the 2010 and
the 2011 ICARUS samples with average energy of $\approx$28~GeV 
represents a total of
$2.1\cdot 10^{12}$ incoming neutrinos. No Cherenkov like event has been
detected in ICARUS. The experimental event rates are compared in
table~\ref{tab:icarates} with the expected ones, for $\delta=0$ and
$\delta=5\cdot 10^{-5}$. The results presented in the table clearly show
that the latter value for $\delta$ can be excluded.
Taking into account both the absence of narrow e$^+$e$^-$
pairs pointing directly to the beam direction and the presence of several
high energy charged current events 
(for instance, 25$\pm$5 events 
with deposited energy in excess of 25~GeV are present in 
Fig.~\ref{fig:ehadr} against 18$\pm$1.3 expected), 
we can set the
limit $\delta < 2.5\cdot 10^{-8}$ at 90\%~CL for CNGS
neutrinos\footnote{$\delta=2.5\cdot 10^{-8}$ corresponds to a 
10\% probability of observing zero events for an exposure of 
$6.70 \cdot 10^{21}$~t$\cdot$pot}, 
comparable to the limit $\delta <
1.4\cdot 10^{-8}$ established by SuperKamiokande~\cite{SUPERKAM} 
from the lack of
depletion of atmospheric neutrinos, and somewhat larger than the lower
energy velocity constraint $\delta < 4\cdot 10^{-9}$ from
SN1987a~\cite{SN1987a}.  
The ICARUS events already collected
during 2011 represent conservatively 
a factor three
higher statistics and should provide more accurate information on the indicated
process. A similar increase in statistics is expected from the 2012 CNGS
run.
However, due to the $\delta^3$ dependence of the pair bremsstrahlung
cross section, no major change of the $\delta$ limit can be expected if no
e$^+$e$^-$ event will be found in the final data sample.
\section*{Conclusions \label{Sect:Concl}}
The spectra and rates at Gran Sasso of neutrino and e$^+$e$^-$ 
for the CNGS beam have been computed in the theoretical framework 
presented in~\cite{cohenPRL,colemanPRD}. In particular, pair bremsstrahlung
events have been accounted for during the propagation of neutrinos from CERN to
Gran~Sasso National Laboratory. The resulting neutrino spectra
and rates for $\delta\approx 5\cdot
10^{-5}$ as suggested by OPERA are significantly different from the
unaffected ones. Preliminary results from the ICARUS experiment do not
support any statistically significant deviation from the unperturbed
spectrum and therefore exclude $\delta$ values comparable to the one
claimed by OPERA.

Furthermore ICARUS did not detect any e$^+$e$^-$ event, despite a few
millions were expected for $\delta=5\cdot 10^{-5}$. The lack of 
e$^+$e$^-$-like event translate into a 90\% CL limit of 
$\delta<2.5 \cdot 10^{-8}$ for multi--GeV neutrinos.
\begin{acknowledgments}
The ICARUS Collaboration acknowledges the fundamental contribution
of INFN to the construction and operation of the experiment.
The Polish groups acknowledge the support of the Ministry of Science and
Higher Education in Poland, including project 637/MOB/2011/0 and
grant number N N202 064936.
The work of A. Cohen was supported by the U.S. Department of Energy
Office of Science.
Finally we thank CERN, in particular the CNGS staff, for the successful 
operation of the neutrino beam facility
\end{acknowledgments}


\begin{thebibliography}{999.}
%
\bibitem{OperaSuper} T.~Adam et al., (The OPERA Collaboration),
arXiv:1109.4897v2, (2011)
%
\bibitem{cohenPRL} A.G.~Cohen, and S.L.~Glashow, Phys. Rev. Lett., 
{\bf 107}, 181803 (2011)
%
\bibitem{SUPERKAM} Super‐Kamiokande
Collaboration,
Y.  Ashie et al., “A Measurement of atmospheric neutrino oscillation
parameters by SUPER-KAMIOKANDE I,” Phys.Rev.  D71 (2005) 112005,
arXiv:hep-ex/0501064 [hep-ex].  Super-Kamiokande Collaboration, S.
Desai et al., “Study of TeV neutrinos with upward showering muons in
Super-Kamiokande,” Astropart.  Phys.  29 (2008) 42–54, arXiv:0711.0053
[hep-ex].  Super-Kamiokande Collaboration, M.  E.  Swanson et al.,
“Search for Diffuse Astrophysical Neutrino Flux Using Ultrahigh Energy
Upward-Going Muons in Super-Kamiokande I,” Astrophys.  J.  652 (2006)
206–215, arXiv:astro-ph/0606126 [astro-ph].
%
\bibitem{SN1987a} M.J.~Longo, ``Tests of Relativity from SN1987a'' Phys. Rev.  
{\bf D36} 3276 (1987) and references therein.
%
\bibitem{ICARUS0} C.~Rubbia, ``The Liquid-Argon Time Projection Chamber:
A New Concept For Neutrino Detector'', CERN-EP/77-08 (1977).
%
\bibitem{ICARUS1} ICARUS Collaboration, F.~Arneodo et al.
  ``Observation of long ionizing tracks with the ICARUS T600 first half-module'',
  NIMA {\bf 508}, 287 (2003)
%
\bibitem{ICARUS2} ICARUS Collaboration, S.~Amerio et al., 
  ``Design, construction and tests of the ICARUS T600 detector'',
  NIMA {\bf 526}, 329 (2004)
%
\bibitem{ICARUS3} ICARUS Collaboration, M.~Antonello et al., 
``Underground operation of the ICARUS
T600 LAr‐TPC: first results'', JINST {\bf 6}, P07011 (2011)
%
\bibitem{CNGSsim1} A.~Ferrari, A.M.~Guglielmi, and P.R.~Sala,
Nuclear Physics B (Proceedings Supplements), {\bf 168}, 169-172 (2007) 
%
\bibitem{CNGSsim2} A.~Ferrari, A.M.~Guglielmi, M.~Lorenzo-Sentis, 
S.~R\"osler, P.R.~Sala,  and L.~Sarchiapone,
  ``An updated Monte Carlo calculation of the CNGS neutrino beam'',
  AB-Note-2006-038, CERN-AB-Note-2006-038 (31 pages), Geneva, CERN,
  20 Aug 2007.  
%
\bibitem{FLUKA1} G.~Battistoni, S.~Muraro, P.R.~Sala, F.~Cerutti, A.~Ferrari,
S.~R\"osler, A.~Fass\`o, J.~Ranft, 
``The FLUKA code: Description and benchmarking'',
Proceedings of the Hadronic Shower Simulation Workshop 2006,
Fermilab 6--8 September 2006, M.Albrow, R. Raja eds., 
AIP Conference Proceeding 896, 31-49, (2007)  
%
\bibitem{FLUKA2} A.~Ferrari, P.R.~Sala, A.~Fass\`o, and J.~Ranft, 
``FLUKA: a multi-particle transport code'', 
CERN-2005-10 (2005), INFN/TC\_05/11, SLAC-R-773 
%
\bibitem{Collaz} G.~Collazuol, A.~Ferrari, A.~Guglielmi, P.R.~Sala,
  Nucl. Instr. Meth. {\bf A449}, 609 (2000)
%
\bibitem{colemanPRD} S.~Coleman, S.L.~Glashow, Phys. Rev. {\bf D59}, 
116008 (1999) 
%
\end{thebibliography}
\end{document}